\documentclass[12pt]{article}
\usepackage{latexsym}
\usepackage{graphics}
\usepackage{color}
\begin{document}

\begin{titlepage}
\begin{flushright}
     {\bf UK/01-04}  \\
\end{flushright}

\begin{center}
{\large\bf Chiral Properties of Pseudoscalar Mesons \\
on a Quenched $20^4$ Lattice with Overlap Fermions}  
  
\vspace{1.5cm}

{S.J. Dong$^{a}$, T. Draper$^{a}$, I. Horv\'{a}th$^{a}$, F.X.
Lee$^{b,c}$, K.F. Liu$^{a}$, and J.B. Zhang$^{d}$}

\bigskip
\bigskip

{\it $^{a}$Dept.\ of Physics and Astronomy, University of Kentucky,
Lexington, KY 40506\\
$^{b}$Center for Nuclear Studies, Dept.\ of Physics, George Washington University,
Washington, DC 20052 \\
$^{c}$Jefferson Lab, 12000 Jefferson Avenue, Newport News, VA 23606 \\
$^{d}$CSSM and Dept.\ of Physics and Math.\ Physics,
University of Adelaide, Adelaide, SA 5005, Australia}
\end{center}
 
 

\vspace{12pt}
 
\begin{abstract}
The chiral properties of the pseudoscalar mesons are studied numerically on a
quenched $20^4$ lattice with the overlap fermion. We elucidate the role of the
zero modes in the meson propagators, particularly that of the pseudoscalar
meson. The non-perturbative renormalization constant $Z_A$ is determined from
the axial Ward identity and is found to be almost independent of the quark mass
for the range of quark masses we study; this implies that the $O(a^2)$ error is
small. The pion decay constant, $f_{\pi}$, is calculated from which we
determine the lattice spacing to be 0.148 fm. We look for quenched chiral log
in the pseudoscalar decay constants and the pseudoscalar masses and we find
clear evidence for its presence.  The chiral log parameter $\delta$ is
determined to be in the range 0.15 -- 0.4 which is consistent with that
predicted from quenched chiral perturbation theory.
\end{abstract}

\vfill
\end{titlepage}

\section{Introduction}

One of the main goals of lattice QCD is to understand from first principles
low-energy phenomenology as a consequence of chiral symmetry.  Recent advances
in the formulation of chiral fermions on the lattice hold great promise for
studying chiral symmetry of QCD at finite lattice spacing~\cite{neu00}.

Neuberger's overlap fermion~\cite{neu98}, derived from the overlap
formalism~\cite{nn95}, is such a chiral fermion on the lattice and has been
implemented numerically to study the chiral
condensate~\cite{ehn99b,hjl99,deg01b,ghr01}, quark mass~\cite{dll00,ghr01},
renormalization constants~\cite{dll00,ghr01,hjl01}, and short-distance current
correlators~\cite{deg01a} and to check chiral symmetry~\cite{dll00,eh00} and
scaling~\cite{dll00}. However these studies are limited to small volumes due to
the large numerical cost associated with approximating the matrix sign
function. In this paper, we shall study physical observables, such as the
pseudoscalar meson masses and pion decay constants, close to the physical $u,d$
quark mass. As such, we need to work on a lattice which is at least 3 times
larger than the Compton wavelength of the pion with the smallest mass in order
to alleviate finite volume effects. We work on a $20^4$ lattice with $a =
0.148\, {\rm fm}$ as determined from the pion decay constant $f_{\pi}$. This
gives the lattice size $La = 3.0\, {\rm fm}$ and the smallest pion mass is
$\sim 280\, {\rm MeV}$. Thus, the lattice size is $\sim 4$ times the Compton
wavelength of the lowest-mass pion.

This paper is organized as follows. We will give the numerical details of the
calculation in Sec.~2. In Sec.~3, we shall discuss the effect of the zero modes
in the meson propagators. In view of the fact that the scalar condensate
receives a contribution from the zero modes (which goes away in the infinite
volume limit) through the generalized Gell-Mann-Oakes-Renner (GOR) relation,
the pseudoscalar correlator should also be contaminated by the zero modes. We
have observed the effect of the zero modes in the pseudoscalar propagator at
small quark mass.  After clarifying the zero mode issue, we proceed to
calculate the non-perturbatively determined renormalization constant $Z_A$ from
the axial Ward identity and the pion decay constant $f_{\pi}$. We find that
$f_{\pi}$ is free of the quenched chiral log singularity and has a small error
in the chiral limit. Thus, it is a good quantity to set the lattice scale. We
present the results in Sec.~4. In Sec.~5, we explain our effort in searching
for the predicted quenched chiral logs. We see the chiral logs in the
pseudoscalar masses, the pseudoscalar matrix element $f_P$, and the
$f_P/f_{\pi}$ ratio at very small quark masses. A summary is given in Sec.~6.

\section{Numerical Details}

For Neuberger's overlap fermion~\cite{neu98}, we adopt the following form for
the massive Dirac operator~\cite{hjl01,afp00,cap00}
\begin{equation}  \label{neu}
D(m_0)= (1 - \frac{m_0a}{2\rho})\rho D(\rho) + m_0a,
\end{equation}
where
\begin{equation}
D(\rho) = 1 + \gamma_5 \epsilon (H),
\end{equation}
so that
\begin{equation}
D(m_0) = \rho + \frac{m_0a}{2} + (\rho - \frac{m_0a}{2} ) \gamma_5 \epsilon (H),
\end{equation}
where $\epsilon (H) = H /\sqrt{H^2}$ is the matrix sign function and $H$ is
taken to be the hermitian Wilson-Dirac operator, i.e. $H = \gamma_5 D_w$.  Here
$D_w$ is the usual Wilson fermion operator, except with a negative mass
parameter $- \rho = 1/2\kappa -4$ in which $\kappa_c < \kappa < 0.25$. We take
$\kappa = 0.19$ in our calculation which corresponds to $\rho = 1.368$. The
massive overlap action is so defined so that the tree-level renormalization of
mass and wavefunction is unity.  The bare mass parameter~\footnote{Note that we
used a different normalization in the action before in \cite{dll00}. As a
result the bare mass here is equal to $\rho$ times the bare mass in
\cite{dll00}.}, $m_0$, is proportional to the quark mass without an additive
constant which we have verified numerically in a previous study~\cite{dll00}.

We adopt the optimal rational approximation~\cite{ehn99a,dll00} to approximate
the matrix sign function. The inversion of the quark matrix involves nested do
loops in this approximation. It is found that it is cost effective to project
out a relatively few eigenmodes with very small eigenvalues in the operator
$H^2$ in order to reduce the condition number and speed up the convergence in
the inner do loop~\cite{ehn99b,dll00}. At the same time, this improves chiral
symmetry relations such as the Gell-Mann-Oakes-Renner relation~\cite{dll00}.
However, it is shown~\cite{ehn99c} that the density of these small eigenmodes
grows as $e^{\sqrt{a}}$ with $a$ being the lattice spacing.  As a result, it is
very costly and impractical to work on large volumes with the lattice spacings
currently used. There are simply too many small eigenmodes to be projected out.

For this reason, we explore other options to clear this hurdle.  We have tested
the tree-level tadpole-improved L\"{u}scher-Weisz gauge action~\cite{lw85} and
find that the density of these small eigenvalue modes is decreased to a point
where it becomes feasible to go to large volumes with a lattice size 4 times
the Compton wavelength of the lightest pion. We further find that the
anisotropic action~\cite{mp97} requires projection of more small eigenvalues in
$H^2$ in order to achieve the same convergence in the inner loop than does the
isotropic one. Thus, we decide to use the isotropic action. We also find that
using the clover action with either sign requires the projection of more small
eigenvalue modes. Therefore we use the Wilson action for $H$ in the Neuberger
operator. On a $20^4$ lattice with $\beta = 7.60$ tree-level tadpole-improved
L\"{u}scher-Weisz gauge action, we project out 85 small eigenmodes.  Beyond
these eigenmodes, the level density becomes large. As a result, the number of
conjugate gradient steps is about 345 for the inner loop and about 300 for the
outer loop. While the number for the inner do loop seems to be fairly
independent of the lattice volume, the number for the outer do loop is about a
factor of 2 larger than those for the Wilson gauge action on small
volumes~\cite{dll00}.

Since the conjugate gradient algorithm accommodates multiple masses with a
minimum overhead, we calculated 16 quark masses ranging from $m_0a = 0.01505$
to $m_0a = 0.2736$ which are listed in Table 1 together with the number of
configurations for each mass.

\begin{table}[b]
\begin{center}
\setlength{\tabcolsep}{1.5mm}
\caption{Quark mass $m_0a$ and number of gauge configurations are listed.}
\bigskip
\begin{tabular}{l|llllllll}
\hline
$m_0a$& 0.01505 & 0.01642 & 0.01915 & 0.02736 & 0.04104 & 0.05472 & 0.06840 & 0.08208 \\
\# cfg. & 25   & 25     & 63      &  63     &  53     & 63      &  25     & 63  \\
\hline
$m_0a$& 0.09576 & 0.1094  & 0.1368  &  0.1642  & 0.1915  & 0.2189  & 0.2462  &  0.2736   \\
\# cfg. & 25  &  63     & 25      & 63       & 25      & 63      & 25      & 63  \\
\hline
\end{tabular}
\end{center}
\end{table}

From the string tension with $\sqrt{\sigma} = 440\, {\rm MeV}$, we find that $a
= 0.13 \,{\rm fm}$. However, as we shall see later in Sec.~4, the scale
determined from $f_{\pi}$ is 0.148(2) fm which makes the physical length of the
lattice to be 3.0 fm. The smallest pion mass turns out to be $\sim 280\,{\rm
MeV}$ so that the size of the lattice is $\sim 4.1$ times of the Compton
wavelength of the lowest mass pion and more than 4 times for the heavier ones.

We adopt the periodic boundary condition for the spatial dimensions and the
fixed boundary condition in the time direction so that we can have effectively
a longer range of time separation between the source and sink to examine the
meson propagators with small quark masses which lead to longer correlation
lengths. The source of the meson interpolation field is placed at the 3rd time
slice and we consider the sink as far as the 16th time slice to mitigate the
boundary effect. This gives us a time separation of 13.

We have varied statistics for different quark masses. Of the 16 cases, 7 have
25 gauge configurations, one has 53, and the remaining 8 have 63
configurations. In order to carry out correlated fits to extrapolate
observables to the physical pion mass, we construct the covariance matrix by
embedding the one with smaller dimension, e.g. 25 and 53 into the one with
dimension 63 in a block diagonal form. For example, the covariance $C_{ij}$ for
the one with 25 configurations is constructed so that $C_{ij} (i \le 25, j >
25) =0, C(i > 25, j \le 25) = 0$ and $C_{ij} ( i,j > 25) = \delta_{ij}$.

\section{Zero Mode Effects in Meson Propagators}

The quark zero mode is known to contribute to the vacuum scalar density
$\langle\bar{\psi}\psi\rangle$ on a finite volume. The latter can be written in
the following form for small quark mass $m_0$\footnote{We shall address the
quenched chiral log issue separately in Sec.~\ref{qcl}.}

\begin{equation}  \label{qcon}
- \langle\bar{\psi}\psi\rangle = \frac{\langle|Q|\rangle}{m_0 V} +
c_0 + c_1 m_0,
\end{equation}
where $Q$ is the topological charge which, according to the Atiya-Singer
theorem, is the difference between the number of left-handed and right-handed
zero modes (i.e. $Q = n_{-} - n_{+}$) and has been shown to hold for overlap
fermions or other local fermion actions which satisfy the Ginsparg-Wilson
relation~\cite{hln98,lus98}. Since $\langle|Q|\rangle$ grows as $\sqrt{V}$, the
zero mode contribution vanishes in the infinite volume limit while keeping
$m_0$ fixed at a non-zero value.  Thus the quark condensate which is the
infinite volume and zero mass limit of $\langle\bar{\psi}\psi\rangle$ is
represented by $c_0$ in Eq. (\ref{qcon}).  However, on a fixed finite volume
lattice, this zero mode contribution is divergent for small enough $m_0$. This
was first observed in the domain-wall formulation~\cite{che98} and is also seen
in the overlap fermion~\cite{ldl00}.  Here we reproduce it in Fig. 1 which
shows the divergent part of $\langle\bar{\psi}\psi\rangle$ from the zero modes
for a $6^3 \times 12$ lattice with the Wilson gauge action at $\beta = 5.7$.

\begin{figure}[tb]
\includegraphics{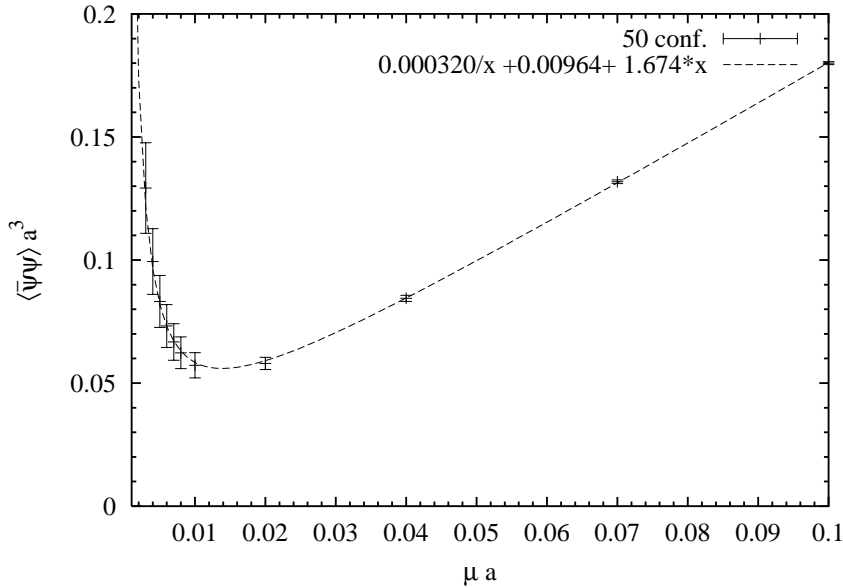}
\vspace{6cm}
\caption{$\langle \bar{\psi}\psi\rangle$ as a function of the
quark mass. We used 50 configurations of a $6^3 \times 12$
lattice with Wilson gauge action at $\beta = 5.7$. Here $\mu a = m_0a/2\rho$.
} \label{z2qw_s50}
\end{figure}

We see from Fig. 1 that $c_0$ is non-zero in this range of the quark mass and
upon extrapolation to the infinite volume before taking the chiral limit
defines the quark condensate $ - \Sigma$. However, if one keeps volume fixed
and lets the quark mass approach zero, e.g.  $m_0a < 0.001$, it is then
found~\cite{ehn99b,hjl99,ldl00} that $c_0$ becomes zero. It is
known~\cite{ls92} that when the size of the lattice is much smaller than the
pion Compton wavelength, i.e. $ L \ll 1/m_{\pi}$, the constant term vanishes
and $\langle\bar{\psi}\psi\rangle$ is proportional to $m_0 \Sigma^2 V$ for
small masses aside from the $\frac{\langle |Q|\rangle}{m_0 V}$ term. Using
finite size scaling, the chiral condensate $\Sigma$ can be
extracted~\cite{hjl99}.

While we have a reasonably good understanding of the role of zero modes in
$\langle\bar{\psi}\psi\rangle$, their role in the hadron propagators is only
beginning to be investigated in the domain-wall formalism~\cite{bcc00} and the
overlap formalism~\cite{deg01b,ghr01} and its detailed influence on the hadron
propagators is not fully understood.  We shall investigate it in the
pseudoscalar meson channel. There has been some concern about the behavior of
the pion mass. It is not clear if it approaches zero in a finite volume in the
chiral limit. In fact, one finds that the pion mass squared appears to approach
a non-zero value on small lattices~\cite{ldl01,dh01}. To understand the
situation, we first examine the generalized Gell-Mann-Oakes-Renner (GOR)
relation

\begin{figure}[t]
\includegraphics{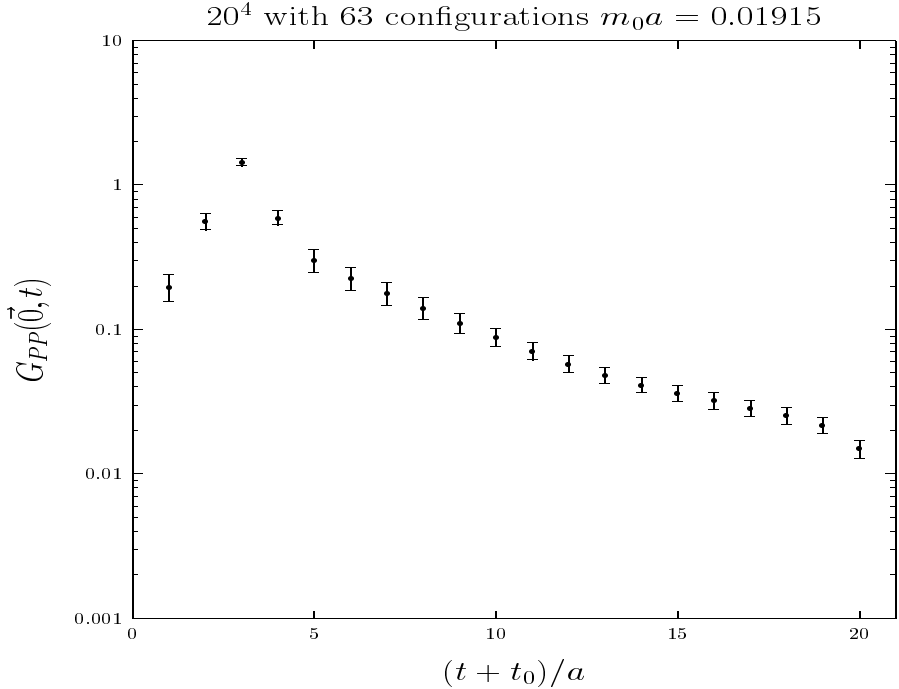}
\vspace{8cm}
\caption{Pion propagator for $m_0a = 0.01915$ for all 63 configurations. The
source is placed at $t_0/a = 3$}
\end{figure}

\begin{equation}   \label{gor}
\frac{2(2m_0)^2}{V}\int d^4x d^4y \langle \pi^a(x) \pi^a(y)\rangle = - 2 m_0
\langle\bar{\psi}\psi\rangle.
\end{equation}

The left hand side of Eq. (\ref{gor}) has a contribution of
$\frac{2\langle|Q|\rangle}{V}$ as does the right hand side. Assuming the
remainder of the pseudoscalar susceptibility is dominated by the pion, it is
approximately $ -f_{\pi}^2 m_{\pi}^2$.  Comparing with Eq. (\ref{qcon}), we get

\begin{equation}  \label{m_pi}
f_{\pi}^2 m_{\pi}^2 = 2m_0 c_0 + 2m_0^2 c_1.
\end{equation}

\begin{figure}[t]
\includegraphics{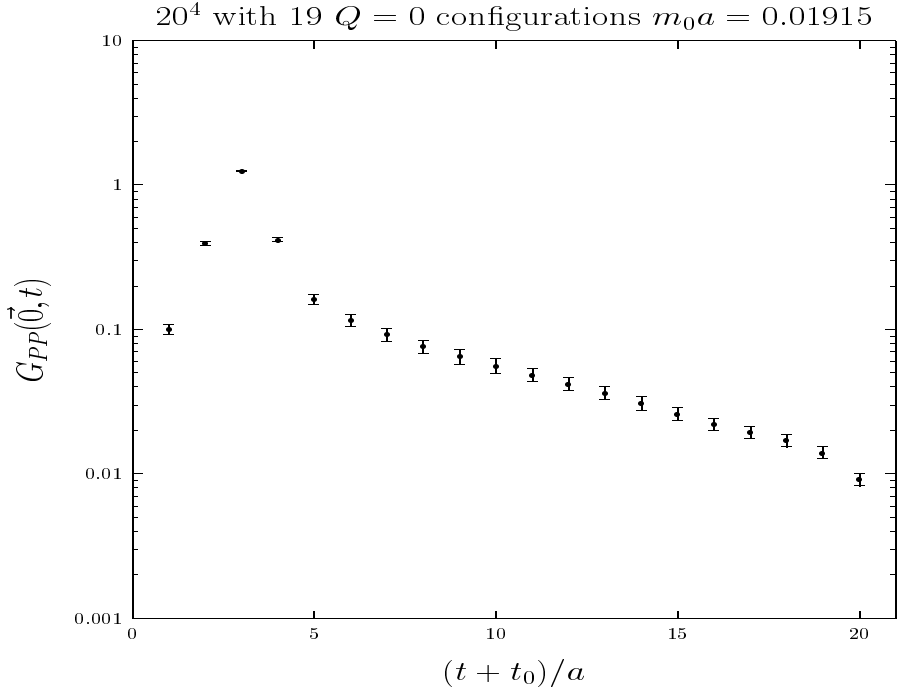}
\vspace{8cm}
\caption{Pion propagator for $m_0a = 0.01914$ for 19 configurations with
$Q = 0$. }
\end{figure}

From this we see that $m_{\pi}^2$ approaches zero as $m_0$ for those volumes
where $c_0$ is non-zero. On the other hand, when the $m_0$ is so small that
$c_0$ becomes zero for certain fixed volume, the r.h.s.\ of Eq. (\ref{m_pi}) is
survived by the next leading term with $c_1$ which is proportional to
$m_0^2$. As a result, one expects that $m_{\pi}$ to be linearly proportional to
$m_0$. Here we have ignored the complication due to the quenched chiral log
which we will address in Sec.~5.

Next, we turn to the zero-momentum pseudoscalar propagator $\int d^3x \langle
\pi^a(x) \pi^a(0)\rangle$ which has two terms due to the zero modes as pointed
out in the study of domain wall fermions~\cite{bcc00}
\begin{eqnarray} \label{pi_propa}
\int d^3x \langle \pi(x) \pi(0)\rangle &=& \int d^3x [\sum_{i,j = zero\, modes}
\frac{tr(\psi^{\dagger}_j(x)\psi_i(x))tr(\psi^{\dagger}_i(0)\psi_j(0))}{m_0^2}
\nonumber \\
 +& 2&\sum_{i = 0, \lambda > 0}\frac{tr(\psi^{\dagger}_{\lambda}(x)\psi_i(x))
tr(\psi^{\dagger}_i(0)\psi_{\lambda}(0))}{m_0 (\lambda^2 + m_0^2)}]
+\frac{|\langle 0|\pi(0)|\pi\rangle|^2 e^{- m_{\pi}t}}{2 m_{\pi}}.\quad
\end{eqnarray}
The first term is purely the zero-mode contribution. The second term is the
cross term between the zero modes and the non-zero modes. We have used the
property that the non-zero modes come in pairs which are related by $\gamma_5$,
i.e. $\gamma_5 \psi_{\lambda} = \psi_{ - \lambda}$. Upon integrating the
propagator with respect to time, we find
\begin{equation} \label{pi_sus}
\int d^4x \langle \pi(x) \pi(0)\rangle
= \frac{\sum_{i = 0} tr(\psi^{\dagger}_i(0)\psi_i(0))}{m_0^2}
+ \frac{\sum_{i =0, \lambda > 0} \delta_{i\lambda}tr(\psi^{\dagger}_i(0)
\psi_{\lambda}(0))}{m_0 (\lambda^2 + m_0^2)} \\
+ \frac{|\langle 0|\pi(0)|\pi\rangle|^2 }{2 m_{\pi}^2}.
\end{equation}
Comparing with the generalized GOR relation in Eq. (\ref{gor}), we see the
first term corresponds to $\frac{\langle|Q|\rangle}{m_0 V}$ term in
Eq. (\ref{qcon}) and the second term vanishes due to the orthogonality between
the zero modes and the non-zero modes. In either case, we expect that the
number of zero modes grows with $\sqrt{V}$ and the eigenfunction $\psi(0)
\propto 1/\sqrt{V}$. As a result, both zero-mode terms in Eqs.
(\ref{pi_propa}) and (\ref{pi_sus}) decrease with volume like $1/\sqrt{V}$ and
are finite volume artifacts.

The most straight-forward way of isolating the zero mode contribution is to
calculate their eigenvectors and project them out as is done in
Ref.~\cite{gpl01}. However, this is very costly, as costly as calculating the
quark propagator itself.

We show the pseudoscalar propagator for a light quark mass ($m_0a = 0.01915$)
in Fig. 2 and we see that there appears a kink at $ t/a \sim 8 - 9$ (The source
is placed at $t_0/a = 3$ so that it appears at $(t + t_0)/a \sim 11- 12$.)  To
further explore its origin, we separate the 63 configurations into 19 with
trivial topology (i.e. $Q = 0$) and 44 with non-trivial topology (i.e.  $Q \neq
0$) and plot the respective pseudoscalar propagators in Figs. 3 and 4.

\begin{figure}[t]
\includegraphics{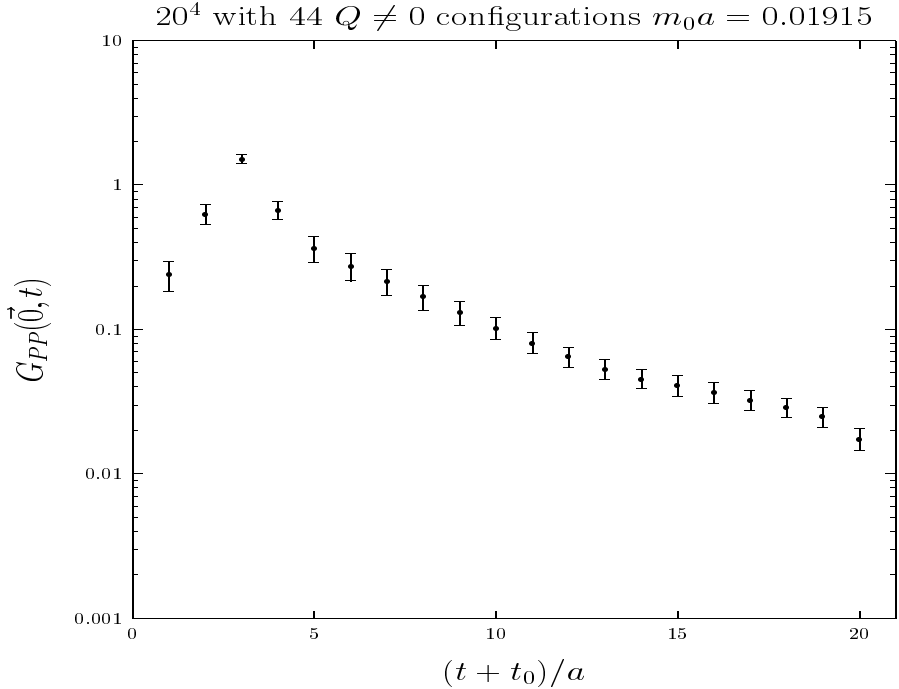}
\vspace{8cm}
\caption{The same as in Fig. 3 for 44 configurations with $Q \neq 0$. }
\end{figure}

We see that the propagator on the $Q = 0$ configurations (Fig. 3) has a single
exponential all the way from $ t/a = 5$ to 14. Upon fitting a single
exponential in this range, we find $m_Pa = 0.153(12)$. On the other hand, the
propagator of the $Q \neq 0$ configurations (Fig. 4) still has a pronounced
kink at $t/a \sim 8$. When we fit it in the range $t/a = 8 - 14$, the mass is
$m_Pa = 0.146(32)$ which is quite consistent with that from the $Q = 0$
configurations. On the other hand, when we fit the time range $t/a$ from 4 to
8, $m_Pa = 0.245(58)$ which is quite a bit higher than that of the $Q = 0$
configurations. We take this as the evidence that the zero-mode contributions,
the combined direct and cross terms, fall off faster in $t/a$ than the
pseudoscalar mass. We can thus use the time separation as the filter to obtain
the masses and decay constants of the physical pseudoscalar mesons. Now we can
understand why in the previous studies of $m_P$, the $m_P^2$ does not approach
zero with the quark mass~\cite{ldl01,dh01}. In Ref.~\cite{ldl01}, the lattice
size is $6^3 \times 12$ with $\beta = 5.7$.  Since it uses the fixed boundary
condition in the time direction, the maximum usable time separation is about
7. This translates into time separation of $\sim 9$ on our lattice.  In
Ref.~\cite{dh01}, the lattice size is $12^3 \times 24$ with $\beta = 5.9$.  In
this study, the authors use the periodic boundary condition in time, time
slices up to 12 are fitted which corresponds to $\sim 10$ on our lattice. In
either case, the fitted time range is expected to be contaminated by the zero
mode contribution and result in a higher mass for a given quark mass. To verify
this, we fit our data in the time range $ t/a = 5 - 9$ and plot the resulting
$m_P^2 a^2$ in Fig. 5. We see that they indeed don't approach zero with a fit
linear and quadratic in $m_0a$ (dashed line), similar to those shown in
Refs.~\cite{ldl01,dh01}. One may attempt to interpret the data to include a
chiral log and force the pion mass to go to zero in the chiral limit.  This can
be misleading. We shall defer the discussion of the complication due to the
quenched chiral log in Sec.~\ref{qcl}.

\begin{figure}[t]
\includegraphics{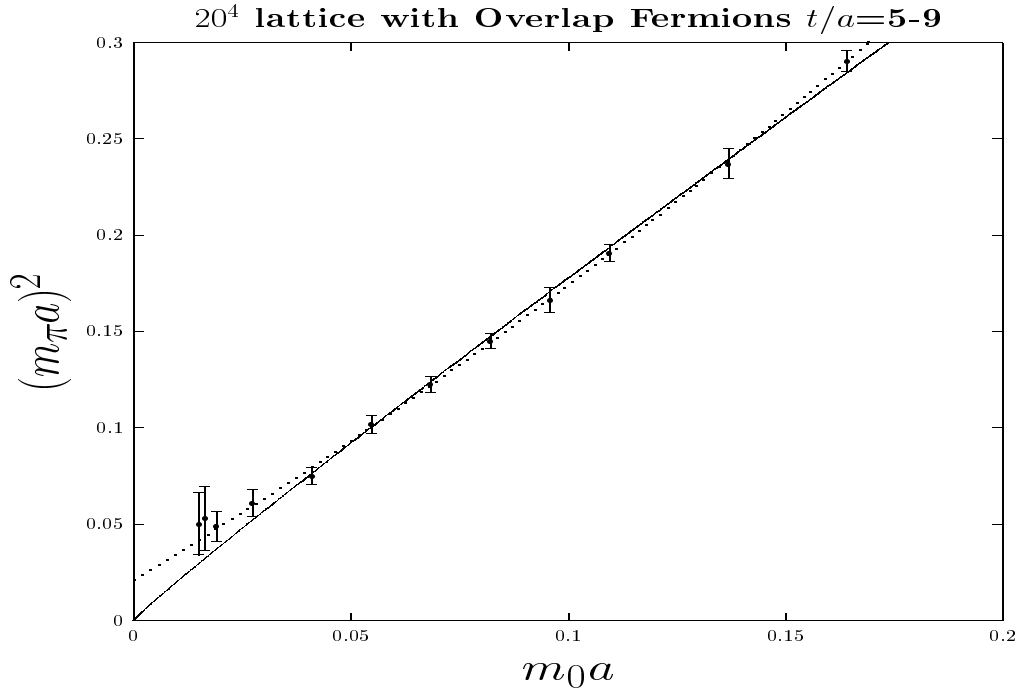}
\vspace{8cm}
\caption{$m_P^2 a^2$ vs $m_0a$ with a fit of $m_Pa$ in the window of
$ t/a = 5 - 9$. The dotted line is a fit linear and quadratic in $m_0a$.
The solid line is a fit including the chiral log.}
\end{figure}

In principle, one can overcome the zero mode problem by fitting the zero
momentum pion propagator $G_{PP}(\vec{p} = 0, t)$ in a time range beyond the
localized zero mode contribution, such as beyond $ (t + t_0)/a > 12$ in
Fig. 2. Unfortunately, our data at larger time slices are tainted by the
reflection from the fixed boundary at $ (t + t_0)/a = 20 $ when the quark mass
is small so that the fit in the limited time window to avoid both the zero mode
and the boundary reflection is unsettling in this case. For this reason, we
examine the propagator $G_{A_4P}(\vec{p} = 0, t)$ instead. Since the zero modes
on one gauge configuration have the same chirality, the pure zero mode
contribution (i.e.\ the direct zero mode contribution which corresponds to the
first term on the right-hand-side of Eq. (\ref{pi_propa})) vanishes. The
cross-term between the zero modes and the non-zero modes does not vanish, but
is expected to be small due to cancellations.  We plot in Fig. 6
$G_{A_4P}(\vec{p} = 0, t)$ for the same quark mass $m_0 a = 0.01915$ and we do
not see a kink in the range $5 < (t + t_0)/a < 15$ as is in $G_{PP} (\vec{p} =
0, t)$ in Fig. 2.  Thus, we shall calculate the pseudoscalar masses from
$G_{A_4P}(\vec{p} = 0, t)$.  The results of $m_P^2 a^2$ so obtained are plotted
in Fig. 7.

Comparing with Fig. 5, we note that $m_P^2 a^2$ in Fig. 7 which are free from
the zero mode contamination are lower than the corresponding ones in Fig. 5 for
small quark masses. How $m_P^2 a^2$ approaches zero in the chiral limit is
complicated by the presence of the quenched chiral log. We shall postpone this
discussion until Sec.~\ref{qcl}, except to mention that the solid line in
Fig. 7 is the fit with the chiral log and the dashed line is the fit with
linear and quadratic terms in $m_0 a$.

\begin{figure}[hb]
\includegraphics{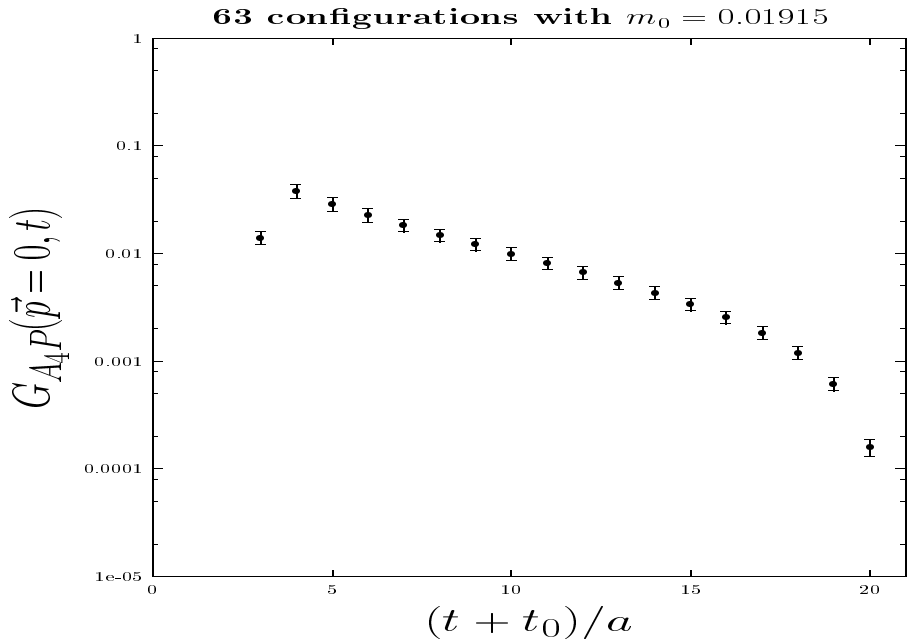}
\vspace{10cm}
\caption{Propagator $G_{A_4P}(\vec{p} = 0, t)$ for $m_0 a = 0.01915$ for all
63 configurations.}
\end{figure}

The zero mode contributions to mesons can be written as
\begin{eqnarray}  \label{zero_meson}
\lefteqn{\int d^3x \langle M(x) M(0)\rangle|_{zero\, modes} = }  \\
& &-  \int d^3x [\sum_{i,j = zero\, modes}
\frac{tr(\psi^{\dagger}_j(x)\gamma_5 \Gamma\psi_i(x))tr(\psi^{\dagger}_i(0)
\bar{\Gamma}\gamma_5\psi_j(0))}{m_0^2} \nonumber \\
& & + 2\sum_{i = 0, \lambda > 0}\frac{tr(\psi^{\dagger}_{\lambda}(x)\gamma_5
\Gamma \psi_i(x))tr(\psi^{\dagger}_i(0)\bar{\Gamma}\gamma_5\psi_{\lambda}(0))}
{m_0 (\lambda^2 + m_0^2)}],
\end{eqnarray}
where $\Gamma$ and $\bar{\Gamma}$ are the gamma matrices for the corresponding
meson interpolation fields. For the pseudoscalar meson as we discussed above,
$\Gamma = - \bar{\Gamma} = \gamma_5$. For the scalar meson (the connected
insertion part), $\Gamma = \bar{\Gamma} = 1$. Since the zero modes are the
eigenstates of $\gamma_5$, i.e. $\gamma_5 \psi_{i = 0} = \pm \psi_{i = 0}$, the
zero mode contribution in the scalar propagator has a negative sign from that
in the pseudoscalar propagator. Thus, it is suggested~\cite{bcc00,deg01b,ghr01}
to consider $\int d^3x [\langle \pi(x) \pi(0)\rangle + \langle \sigma(x)
\sigma(0)\rangle]$ where the zero-mode contributions cancel and at large $t/a$
it should be dominated by the pseudoscalar. However, there is a problem.
\begin{figure}[t]   \label{pi_sq}
\includegraphics{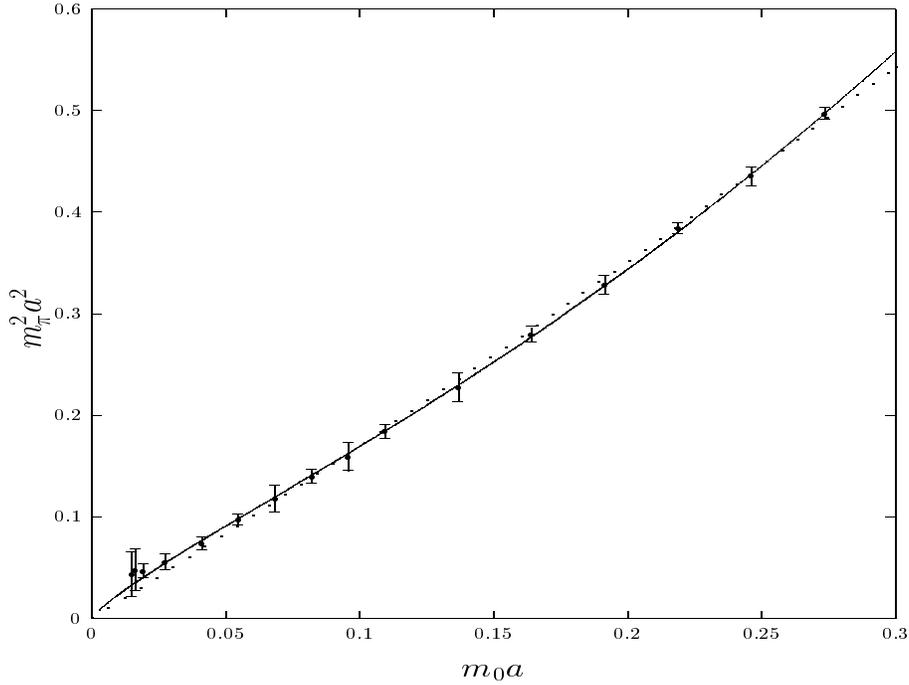}
\vspace{7cm}
\caption{$m_P^2 a^2$ vs $m_0a$ calculated from $G_{A_4P}(\vec{p} = 0, t)$
The linear plus quadratic fit (dotted line) and the
chiral log fit (solid line) from Eq. (\ref{chi_log}) with $\Lambda_{\chi}
= 0.8\, {\rm GeV}$ are discussed in Sec.~\ref{qcl}.}
\end{figure}
As will be shown in a separate publication~\cite{ddl01c}, the large time part
of the isovector-scalar propagator for the quark mass range that we are
concerned turns out to be negative. It is pointed out in a study with
pole-shifting in the Wilson action~\cite{bde01} that it is dominated by the
would-be $\eta'$ and $\pi$ intermediate state which is negative due to
quenching. In the intermediate time region, there is still the contribution
from $\eta'$ and $\pi$ intermediate state besides the $a_0$. As a consequence,
the addition of the pseudoscalar and scalar meson is not a viable solution to
obtaining the pseudoscalar mass.  It is also suggested that the axial-vector
interpolation field with $\Gamma = \bar{\Gamma} = \gamma_4\gamma_5$ does not
have the direct term contribution from the zero modes since all the zero modes
have the same chirality in a given gauge configuration. However, the second
term in Eq. (\ref{zero_meson}) may still have a contribution. Since it is a
cross term, it might be small due to cancellations. Unfortunately, our data on
the $\langle A_4 A_4 \rangle$ correlator are much noisier than the $\langle \pi
\pi \rangle$ correlator, since $\langle 0|A_4|\pi(0)\rangle = \sqrt{2} m_{\pi}
f_{\pi}$ which diminishes when the pion mass is small. We cannot conclude
anything from them.

It appears that, short of projecting out the zero modes from the meson
propagators, fitting the pseudoscalar mass from the propagator
$G_{A_4P}(\vec{p} = 0, t)$ is probably the only practical way of getting
reasonably reliable and accurate pseudoscalar masses for our lattice with fixed
boundary condition.

\section{$Z_A$ and Pion Decay Constant $f_{\pi}$}   \label{Z_A}

It has been pointed out in our earlier work~\cite{dll00} that the
renormalization constant $Z_A$ for the axial current
$A_{\mu}=\bar{\psi}(i\gamma_{\mu}\gamma_5(1 - D/2\rho)\frac{\tau^a}{2})\psi$
can be obtained directly through the axial Ward identity
\begin{equation}  \label{awi}
Z_A\partial_{\mu} A_{\mu} = 2 Z_m m_0 Z_P P,
\end{equation}
where $P =\bar{\psi}(i\gamma_5(1 - D/2\rho)\frac{\tau^a}{2})\psi$ is the
pseudoscalar density.  Since $Z_m = Z_S^{-1}$ and $Z_S = Z_P$ due to the fact
that the scalar density $\bar{\psi}(1 - D/2\rho)\frac{\tau^a}{2}\psi$ and the
pseudoscalar density $P$ are in the same chiral multiplet, $Z_m$ and $Z_P$
cancel in Eq. (\ref{awi}) and one can determine $Z_A$ to $O(a^2)$
non-perturbatively from the axial Ward identity using the bare mass $m_0$ and
bare operator $P$.  To obtain $Z_A$, we shall consider the on-shell matrix
elements between the vacuum and the zero-momentum pion state for the axial Ward
identity

\begin{equation}
Z_A \langle 0|\partial_{\mu} A_{\mu}|\pi(\vec{p} = 0)\rangle
= 2 m_0 \langle 0|P|\pi(\vec{p} = 0)\rangle,
\end{equation}
where the matrix elements can be obtained from the zero-momentum correlators
\begin{eqnarray}
G_{\partial_4A_4P}(\vec{p}= 0, t)& =& \langle\sum_{\vec{x}}\partial_4 A_4(x)
P(0)\rangle \nonumber \\
G_{PP}(\vec{p}= 0, t) & =& \langle\sum_{\vec{x}} P(x) P(0)\rangle.
\end{eqnarray}
The non-perturbative $Z_A$ is then
\begin{equation}   \label{ZA}
Z_A = \lim_{t \longrightarrow \infty} \frac{2 m_0 G_{PP}(\vec{p}= 0, t)}
{G_{\partial_4A_4P}(\vec{p}= 0, t)}.
\end{equation}

\begin{figure}[tb]
\includegraphics{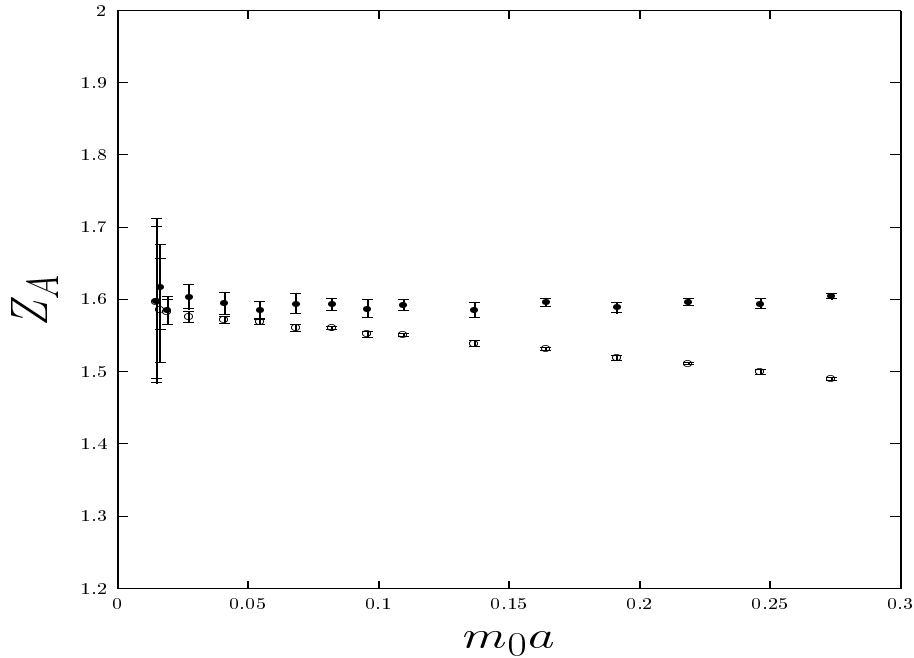}
\vspace{7cm}
\caption{$Z_A$ vs quark mass $m_0a$. The results from Eq. (\ref{ZA}) are
indicated by $\circ$ and those from Eq. (\ref{ZA1}) are indicated by
$\bullet$.}
\end{figure}

There has been extensive study~\cite{hmp91,lss96,lw96,lss97} of the $O(a)$
improvement of the Wilson action and composite operators in relation to the
chiral symmetry and axial Ward identity.  It has been shown~\cite{lss96} that
in the improved mass-independent renormalization scheme, the renormalized
improved axial current and pseudoscalar density have the following form from an
$O(a)$ improved action
\begin{eqnarray}
A_{\mu}^R &=& Z_A (1 + b_A m_q a) \{A_{\mu} + c_A a \partial_{\mu}P\},
\nonumber \\
P^R &=& Z_P (1 + b_P m_q a) P,
\end{eqnarray}
where $m_q = m_0 - m_c$ is the subtracted quark mass and $c_A, b_A$ and $b_P$
are improvement coefficients. The renormalization constants $Z_A$ and $Z_P$ are
functions of the modified coupling $\tilde{g_0}^2 = g_0^2 (1 + b_g m_q a)$.
Now that the overlap fermion is $O(a)$ improved~\cite{knn97}, it satisfies the
Ginsparg-Wilson relation~\cite{neu98b} and the quark mass is not additively
renormalized. As a result, $b_A = c_A = b_P = 0$. The renormalization constant
$Z_A$ determined from Eq. (\ref{ZA}) has only $O(a^2)$ error as a
consequence. $Z_A$ thus computed for all the quark masses are plotted in Fig. 8
as indicated by the open circles. We see that there is still an appreciable
$O(a^2)$ error. Given that the time derivative itself in
$G_{\partial_4A_4P}(\vec{p}= 0, t)$ invokes an $O(a^2)$ error, it would be
better to adopt a definition for $Z_A$ which is devoid of this superfluous
$O(a^2)$ error. This can be achieved by noticing that, at large $t$ where the
pion state dominates the propagator $G_{\partial_4A_4P}(\vec{p}= 0, t)$, one
can effectively make the substitution
\begin{equation}   \label{subst}
G_{\partial_4A_4P}(\vec{p}= 0, t) {}_{\stackrel{\longrightarrow}{t \rightarrow
\infty}} m_{\pi}  G_{A_4P}(\vec{p}= 0, t).
\end{equation}
Consequently, Eq. (\ref{ZA}) becomes
\begin{equation}  \label{ZA1}
Z_A = \lim_{t \longrightarrow \infty} \frac{2 m_0 G_{PP}(\vec{p}= 0, t)}
{m_{\pi} G_{A_4P}(\vec{p}= 0, t)}.
\end{equation}

\begin{table}[htb]
\begin{center}
\caption{Renormalization constant $Z_A$ from Eq. (\ref{ZA}) (column 2),
and Eq. (\ref{ZA1}) (column 3).}
\begin{tabular}{lll}
\hline
$m_0a$ & $Z_A$ (Eq. (\ref{ZA}))& $Z_A$ (Eq. (\ref{ZA1})) \\
\hline
0.2736  & 1.490(2) & 1.605(4)      \\
0.2462  & 1.499(3) & 1.594(6)      \\
0.2189  & 1.511(2) & 1.597(4)      \\
0.1915  & 1.519(3) & 1.589(8)      \\
0.1642  & 1.532(2) & 1.596(5)      \\
0.1368  & 1.539(4) & 1.585(11)     \\
0.1094  & 1.551(2) & 1.592(8)      \\
0.09576 & 1.552(4) & 1.587(12)     \\
0.08208 & 1.560(2) & 1.593(8)      \\
0.06840 & 1.561(5) & 1.594(13)     \\
0.05472 & 1.569(3) & 1.586(12)     \\
0.04104 & 1.572(5) & 1.595(15)     \\
0.02736 & 1.576(8) & 1.604(17)     \\
0.01915 & 1.583(17)  & 1.585(20)   \\
0.01642 & 1.585(72)  & 1.618(60)   \\
0.01505 & 1.596(105) & 1.598(114)  \\
\hline
\end{tabular}
\end{center}
\end{table}

We plot the results of $Z_A$ from Eq. (\ref{ZA1}) in Fig. 8.  Save for the last
two points at the smallest masses ($m_0a = 0.01505$ and 0.01642), the errors
are small. We should point out that it is conspicuously flat as a function of
$m_0a$ indicating that the $O(a^2)$ error from the action and the operators is
small. We fit them in the form $Z_A + b m_0^2 a^2$ and found that $Z_A =
1.589(4), b = 0.175(73)$ with $\chi^2/DF = 0.30 $. We see that $Z_A$ is
determined to the precision of a fraction of 1\%.  We find it to be larger than
the perturbative calculation~\cite{afp00} which gives $Z_A (\mu = 1/a) = 1.213$
for Wilson gauge action with $\beta = 5.85$ which has about the same string
tension as our gauge configurations. $Z_A$ from both Eqs. (\ref{ZA}) and
(\ref{ZA1}) are tabulated in Table 2.

For the pion decay constant, one can look at the ratio of the zero-momentum
correlator
\begin{equation}
G_{A_4P}(\vec{p}= 0, t) = \langle\sum_{\vec{x}} A_4(x) P(0)\rangle,
\end{equation}
and $G_{PP}(\vec{p}= 0, t)$,
\begin{equation}
f_{\pi}a = \lim_{t \longrightarrow \infty}\frac{Z_A G_{A_4P}(\vec{p}= 0, t)}
{\sqrt{m_{\pi}a\, G_{PP}(\vec{p}= 0, t)}} e^{m_{\pi}t/2}.
\end{equation}
For our definition of the isovector axial and pseudoscalar currents, the
experimental $f_{\pi}$ is 92.4 MeV.

Combining with $Z_A$ from Eq. (\ref{ZA}), we obtain
\begin{equation} \label{f_pi}
f_{\pi}a = \lim_{t \longrightarrow \infty}\frac{2 m_0a\, \sqrt{G_{PP}(\vec{p}= 0, t)}
G_{A_4P}(\vec{p}= 0, t)}{\sqrt{m_{\pi}a}\,G_{\partial_tA_4P}(\vec{p}= 0, t)}
e^{m_{\pi}t/2}.
\end{equation}

We first fit the pion masses from $G_{A_4P}(\vec{p}= 0, t)$ and feed them into
Eq. (\ref{f_pi}). Unlike the case with $Z_A$ in Eqs. (\ref{ZA}) and (\ref{ZA1})
where the boundary effects on the common source for the interpolation field $P$
in the numerator and denominator cancel, they don't cancel in $f_{\pi}$. To
correct for this, we replace $\sqrt{G_{PP}(\vec{p}= 0, t)}$ with
$G_{PP}(\vec{p}= 0, t)\,e^{m_{\pi}t/2} /\sqrt{G_{PP}(\vec{p} = 0,\,(N_t +1) a -
t_0)\,e^{m_{\pi}((N_t +1) a - t_0)}}$ which should get rid of the boundary
effect which affects the matrix element $\langle 0| P|\pi\rangle$ associated
with the source in $G_{PP}$. The errors of the ratio are obtained with the
jackknife method. We plot the result in Fig. 9 as a function of $m_0a$ as open
circles. We also tabulate them in Table 3 together with the pion mass computed
from $G_{A_4P}(\vec{p}= 0, t)$.

\begin{figure}[th]
\includegraphics{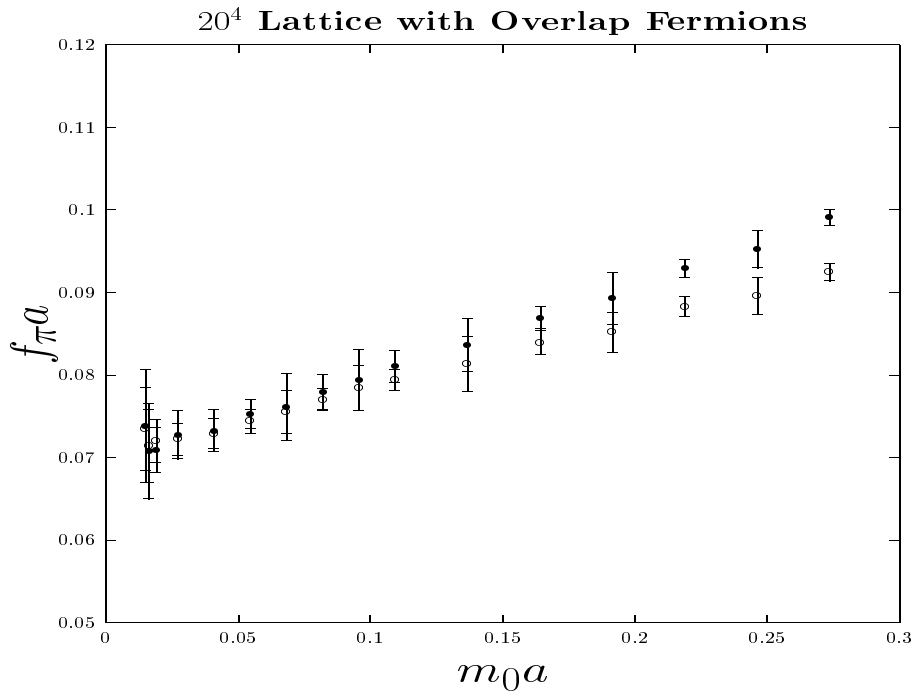}
\vspace{10cm}
\caption{Renormalized $f_{\pi}a$ vs quark mass $m_0a$. }
\end{figure}

As in the case of $Z_A$ defined from $G_{\partial_4A_4P}(\vec{p}= 0, t)$,
$f_{\pi}$ defined from Eq. (\ref{f_pi}) contains $O(a^2)$ error invoked by the
time derivative.  We shall again make the substitution in Eq. (\ref{subst}) and
arrive at
\begin{equation}   \label{f_pi1}
f_{\pi}a = \lim_{t \longrightarrow \infty}\frac{2 m_0a\, \sqrt{G_{PP}(\vec{p}= 0, t)
\,m_{\pi}a} e^{m_{\pi}t/2}}{(m_{\pi}a)^2}.
\end{equation}

The results are listed in Table 3 and plotted in Fig. 9 as $\bullet$. The
difference from the $\circ$ reflects the $O(a^2)$ error due to the time
derivative in $G_{\partial_4A_4P}(\vec{p}= 0, t)$.  With all 16 data points in
a linear fit, we find
\begin{equation}
f_{\pi}a = 0.0691(11) + 0.109(56)\,m_0a,
\end{equation}
with $\chi^2/DF = 0.03$. We note that it yields an error of only 1.6\% in the
chiral limit. Given that it is a physical observable from the fermion operator
without the chiral log in a quenched theory~\cite{sha92} and with such high
precision, it is an ideal quantity to set the scale of the lattice. Comparing
with the experimental value $f_{\pi} = 92.4\, {\rm MeV}$, we determine the
scale of our lattice to be $a = 0.148(2)\, {\rm fm}$. This is about 14\% higher
than that determined from the string tension with $\sqrt{\sigma} = 440\,{\rm
MeV}$ and 20\% larger than that determined from $r_0$. This is quite consistent
with other quenched calculations.

Our results on $f_{\pi}$ and its quark mass dependence are consistent with
those calculated with the domain-wall fermion~\cite{bcc00} and the overlap
fermion~\cite{ghr01} on smaller lattices and higher pion masses.

\begin{table}[htb]
\begin{center}
\caption{Pion masses obtained from $G_{A_4P}(\vec{p}= 0, t)$ are listed
together with $f_{\pi}a$ from Eq. (\ref{f_pi}) (column 4),
and Eq. (\ref{f_pi1}) (column 5), and $f_P a^2$ from Eq. (\ref{f_P_G_PP}).}
\bigskip
\bigskip
\begin{tabular}{llllll}
\hline
$m_0a$ & $m_{\pi} a$ & $m_{\pi}^2 a^2$ & $f_{\pi}a$ (Eq. (\ref{f_pi})) &
 $f_{\pi}a$ (Eq. (\ref{f_pi1})) & $f_P a^2$  \\
\hline
0.2736  & 0.705(4)  & 0.497(6)  & 0.0925(10) & 0.0991(10) & 0.1273(13)  \\
0.2462  & 0.660(7)  & 0.436(9)  & 0.0896(23) & 0.0952(23) & 0.1192(28) \\
0.2189  & 0.620(4)  & 0.384(5)  & 0.0883(12) & 0.0929(11) & 0.1155(14) \\
0.1915  & 0.573(8)  & 0.328(9)  & 0.0852(24) & 0.0893(31) & 0.1082(38)   \\
0.1642  & 0.529(7)  & 0.280(7)  & 0.0839(14) & 0.0869(13) & 0.1048(16) \\
0.1368  & 0.477(15) & 0.228(14) & 0.0813(33) & 0.0837(33) & 0.0983(39) \\
0.1094  & 0.429(8)  & 0.184(7)  & 0.0794(13) & 0.0810(19) & 0.0963(22)  \\
0.09576 & 0.399(17) & 0.159(14) & 0.0784(28) & 0.0794(37) & 0.0934(43)  \\
0.08208 & 0.374(9)  & 0.140(7)  & 0.0770(13) & 0.0779(21) & 0.0939(25)  \\
0.06840 & 0.343(19) & 0.118(13) & 0.0755(26) & 0.0762(40) & 0.0927(49)  \\
0.05472 & 0.312(9)  & 0.0973(56)& 0.0744(15) & 0.0753(18) & 0.0942(30)  \\
0.04104 & 0.272(12) & 0.0740(65)& 0.0729(18) & 0.0733(25) & 0.0933(32)  \\
0.02736 & 0.236(17) & 0.0557(80)& 0.0722(19) & 0.0728(29) & 0.1055(34)  \\
0.01915 & 0.216(16) & 0.0467(69)& 0.0721(26) & 0.0709(27) & 0.1222(42) \\
0.01642 & 0.218(47) & 0.0475(205)&0.0714(44) & 0.0708(58) & 0.145(12) \\
0.01505 & 0.209(53) & 0.0437(221)&0.0734(50) & 0.0738(68) & 0.152(14) \\
\hline
\end{tabular}
\end{center}
\end{table}

\section{Quenched Chiral Logs}  \label{qcl}

In the quenched approximation of QCD, one ignores the virtual quark loops. One
of the consequences is that the flavor-singlet meson propagator (which we shall
refer to as $\eta'$ even though the physical $\eta'$ is not completely
flavor-singlet and has octet mixture) has a double pole in the Veneziano model
for the U(1) anomaly~\cite{ven79,liu92}. As such, it does not move the mass of
the would-be Goldstone boson to the large $\eta'$ mass.  Consequently, this
leads to infrared singular $\eta'$ loops with the hair-pin type diagrams in the
renormalization of hadron masses and certain matrix elements and thus alters
their chiral behaviors from those of full QCD with dynamical fermions.

The first study of the anomalous chiral behavior was done by
Sharpe~\cite{sha92} and Bernard and Golterman~\cite{bg92} in quenched chiral
perturbation theory. They predicted the chiral-log pathologies in the
pseudoscalar meson masses, $\langle \bar{\psi}\psi\rangle$, the $f_K/f_{\pi}$
ratio, etc. The first evidence of the chiral-log was observed by the CP-PACS
Collaboration~\cite{abb00} in the ratio of pseudoscalar meson masses with two
unequal quark masses with the Wilson fermion.  They obtain the chiral log
parameter $\delta = 0.8 - 1.2$. A more extensive study~\cite{bde00} which
invokes the shifting of the real poles of the quark propagator to improve the
otherwise poor chiral properties of the Wilson fermion near the chiral limit
was carried out to examine the quenched chiral logs in the pseudoscalar masses
and the pseudoscalar decay constants, obtaining consistent results with $\delta
= 0.065 \pm 0.013$.  The same quenched chiral log is also observed in the ratio
of $m_{\pi}^2 /m_q$ with the Kogut-Susskind action~\cite{bbo01} with $\delta =
0.061 \pm 0.03$.  All of them are small compared to that expected from the
coupling of the would-be Goldstone bosons (or quark loops in the pseudoscalar
channel) which is responsible for the $\eta'$ mass of the U(1)
anomaly. Recently, the small nonzero eigenvalues of the Overlap Dirac operator
were calculated in a deconfined phase~\cite{kn01} and it was found that the
chiral condensate diverges at the infinite volume limit indicating a quenched
singularity consistent with the quenched chiral perturbation prediction
$\langle\bar{\psi}\psi\rangle \propto m^{\frac{-\delta}{1 + \delta}}$.

Given that the overlap fermion has the promise of exact chiral symmetry on
the lattice, it is natural to look for these chiral singularities and check
if the quenched chiral logs seen in the Wilson and Kogut-Susskind fermions can
be verified with the overlap action. The first attempt to extract the chiral
log from the pion mass on several small volumes was inconclusive~\cite{dll00}.
We shall examine them here on a much larger volume.

The behavior of the quenched chiral logs can be seen from the sigma
model~\cite{bde00} with $U(3) \times U(3)$ where the pseudoscalar field is
represented by
\begin{equation}
U = e^{\phi_0/f} e^{i \sum_{a =1}^8 \lambda_a \phi_a/f},
\end{equation}
where $\lambda_a$ is the SU(3) flavor matrix and $\phi_a$ are the octet
Goldstone boson fields. The U(1) part is described by $\eta'$ field with
$\phi_0 = \sqrt{2/N_f}\eta'/f$ where $N_f$ is the number of flavors which is 3
in our case. The effect of the chiral logs can be understood as the
renormalization effect of integrating out the $\eta'$~\cite{bde00}. The
resulting $SU(3) \times SU(3)$ will be represented by the renormalized $U$
\begin{equation}   \label{U}
U =  e^{-\langle \phi_0^2\rangle/2f^2} e^{\sum_{a = 1}^8 \lambda_a \phi_a/f}.
\end{equation}
In the quenched approximation, the integral representing the $\eta'$ loop
involves only the hair-pin diagram of two would-be singlet Goldstone boson (we
shall refer to it as $\eta$ and it has the same mass as $\pi$) propagators
\begin{equation}
\langle \phi_0^2\rangle = \frac{2}{V N_f} \int d^4x \langle \eta(x) \eta(x)\rangle
= \int \frac{d^4p}{(2\pi)^4 N_f} \frac{- \bar{m_0}^2}{p^2 + m_{\pi}^2}
= \frac{- \bar{m_0}^2}{16 \pi^2 N_f}(\ln \frac{\lambda^2}{m_{\pi}^2} - 1)
\end{equation}
Therefore, the infrared singular part of U in Eq. (\ref{U}) can be represented
by $\delta$
\begin{equation} \label{U_delta}
U = e^{-\delta \ln m_{\pi}^2} e^{\sum_{a = 1}^8 \lambda_a \phi_a/f}
  = (\frac{1}{m_{\pi}^2})^{\delta} e^{\sum_{a = 1}^8 \lambda_a \phi_a/f},
\end{equation}
where
\begin{equation}  \label{delta}
\delta = \frac{\bar{m_0}^2}{16 \pi^2 N_f f^2}.
\end{equation}
From the Witten-Veneziano model of the $\eta'$ mass, $\bar{m_0} \sim 870 {\rm
MeV}$.  This gives an estimate of $\delta = 0.183$.

To see the effects on various physical quantities, one can first look at
pseudoscalar density and axial current operators~\cite{bde00}. In the sigma
model,
\begin{eqnarray}
\bar{\psi} i \gamma_5 \psi & \propto & U - U^{\dagger} \nonumber \\
\bar{\psi} i \gamma_{\mu}\gamma_5 \psi & \propto & i[U^{-1}\partial_{\mu} U
- (\partial_{\mu} U^{-1}) U].
\end{eqnarray}
With U given in Eq. (\ref{U_delta}), one arrives at
\begin{eqnarray}  \label{f_P}
f_P &= &\langle 0|\bar{\psi} i \gamma_5 \psi |\pi(\vec{p} = 0)\rangle
     = (\frac{1}{m_{\pi}^2})^{\delta} \tilde{f_P} \nonumber \\
f_{\pi} &= &\langle 0|\bar{\psi} i \gamma_{\mu}\gamma_5 \psi |\pi(\vec{p} = 0)
\rangle/\sqrt{2}m_{\pi} = \tilde{f_{\pi}}
\end{eqnarray}
where $f_P$ is the pseudoscalar decay constant and $f_{\pi}$ is the axial decay
constant and $\tilde{f_P}$ and $\tilde{f_{\pi}}$ are constants for small quark
masses. Thus, one expects that that $f_P$ is singular as the quark masses
approaches zero in the quenched approximation; whereas $f_{\pi}$ remains a
constant. From the axial Ward identity in Eq. (\ref{awi}) and Eq. (\ref{f_P}),
one expects
\begin{equation}
m_{\pi}^2 \propto m_q \frac{f_P}{f_{\pi}} \propto m_q
(\frac{1}{m_{\pi}^2})^{\delta},
\end{equation}
where $m_q$ is the quark mass and therefore
\begin{equation}  \label{m_pi_delta}
m_{\pi}^2 \propto m_q^{\frac{1}{1 + \delta}}
\end{equation}
which is the behavior predicted in quenched $\chi PT$~\cite{sha92}.

Given $m_{\pi}a$ fitted from $G_{A_4P}(\vec{p}= 0, t)$ and listed in Table 3,
we first fit them in the form in Eq. (\ref{m_pi_delta}) for a range of small
quark mass points. We list the fitted $\delta$ and $\chi^2/DF$ in Table 4.  We
find that $\delta$ ranges from $0.145(66)$ to $0.24(12)$. Beyond this range of
8 to 11 smallest quark masses, the errors are greater than half of the fitted
$\delta$ value.

\begin{table}[htb]
\begin{center}
\caption{Quenched chiral log parameter $\delta$ and $\chi^2/DF$ as fitted
from $m_{\pi}^2 a^2$ in Eq. (\ref{m_pi_delta}).}
\bigskip
\bigskip
\begin{tabular}{lll}
\hline
\# of smallest $m_0a$ & $\delta$ & $\chi^2/DF$  \\
\hline
  8     &  0.24 (12) & 0.24     \\
  9     &  0.224(91) & 0.22     \\
  10    &  0.163(75) & 0.28     \\
  11    &  0.145(66) & 0.28     \\
\hline
\end{tabular}
\end{center}
\end{table}

Since the form in Eq. (\ref{m_pi_delta}) is limited to small $m_0 a$, we shall
also fit $m_{\pi}^2 a^2$ with the form~\cite{bg92,abb00}
\begin{equation} \label{chi_log}
m_{\pi}^2 a^2 = A m_{0}a \{1 -\delta [\ln(Am_0 a/\Lambda_{\chi}^2 a^2) +1]\}
+ B m_0^2 a^2,
\end{equation}
which allows a fit to cover the whole range of 16 quark masses.  The best fits
which give stable values of A and B and with errors less than half of the
fitted values of $\delta$ for a range of $\Lambda_{\chi} = 0.6$ GeV to 1.4 GeV
are listed in Table 5.

\begin{table}[hbt]
\begin{center}
\caption{Quenched chiral log parameter $\delta$ and $\chi^2/DF$ as fitted
from $m_{\pi}^2 $ in Eq. (\ref{chi_log}).}
\bigskip
\bigskip
\begin{tabular}{lllll}
\hline
$\Lambda_{\chi}$ (GeV) & A  & B & $\delta$ & $\chi^2/DF$  \\
\hline
  0.6   &  1.72(7)  & 3.0(8) & 0.23(7)  & 0.18   \\
  0.8   &  1.42(11) & 3.0(8) & 0.28(11) & 0.18     \\
  1.0   &  1.17(24) & 3.0(8) & 0.34(17) & 0.18    \\
\hline
\end{tabular}
\end{center}
\end{table}

They are consistent with those fitted in the exponential form, albeit with
values of $\delta$ between 0.2 and 0.3 which are somewhat higher than those
from the exponential form.

We plot the fit with $\Lambda_{\chi} = 0.8\, {\rm GeV}$ as a solid line in
Fig. 7. To check if there is indeed a bona fide quenched chiral log, we fit
$m_{\pi}^2 a^2$ alternatively with powers in $m_0a$ up to cubic terms and found
\begin{eqnarray}
m_{\pi}^2 a^2 &=& 1.65(5)\, m_0a + 0.54(21)\, m_0^2 a^2,\,\,\,\,\,\,\,\,
\chi^2/DF = 0.90, \nonumber \\
m_{\pi}^2 a^2 &=& 1.88(10)\, m_0a - 2.3(11)\, m_0^2 a^2 + 7.8(30)\, m_0^3 a^3,
\,\,\,\,\,\chi^2/DF = 0.46.
\end{eqnarray}

The fit including the cubic term in $m_0a$ is shown in Fig. 7 as the dotted
line. We see that in either case, the $\chi^2/DF$ is larger than that with the
quenched chiral log in Table 5 which is 0.18. From this, we conclude that the
quenched chiral is seen in our data for $m_{\pi}^2 a^2$.

We should point out that if we were to use the $m_P^2 a^2$ data from the
shorter time range such as the one with $t/a = 5 - 9$ as plotted in Fig. 5 and
insist on fitting them with a power form in Eq. (\ref{m_pi_delta}), we would
find that one can fit the last 12 mass points with $\delta = 0.054 \pm 0.028$
and $\chi^2/DF = 0.98$. This is plotted as the solid line in Fig. 5.  This
serves as a caveat to show that one can misconstrue the contamination of the
zero modes as the quenched chiral log.

\begin{figure}[tb]
\includegraphics{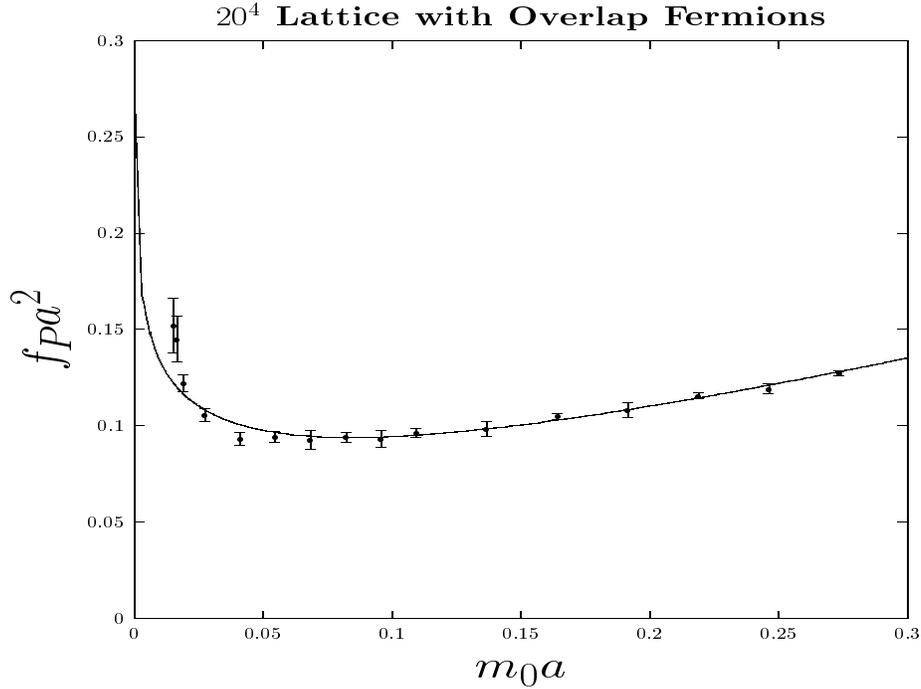}
\vspace{8cm}
\caption{Renormalized $f_Pa^2$ vs quark mass $m_0a$. The solid line is a fit
excluding the two smallest quark masses with $\Lambda_{\chi}$ = 0.8 GeV in
Eq. (\ref{f_P_log}).}
\end{figure}

Finally, we look for the chiral log in $f_P$, which according to
Eq. (\ref{f_P}) should grow in the form of $(\frac{1}{m_{\pi}^2})^{\delta}$.
We calculate the renormalized $f_P$ from
\begin{equation}  \label{f_P_G_PP}
f_P a^2 = \lim_{t/a \gg 1} Z_P \sqrt{G_{PP}(t) 2 m_P a} e^{m_Pt/2},
\end{equation}
where we insert the fitted $m_Pa$ to obtain $f_P a^2$. The $\sqrt{G_{PP}(t)}$
is understood to have included the correction of the boundary reflection
discussed in Sec.~\ref{Z_A} in association with the determination of $f_{\pi}$.
The renormalization constant $Z_P$ is the same as the scalar renormalization
constant $Z_S$. The latter is determined from the matching of renormalization
group invariant quark masses at fixed pseudoscalar mass~\cite{hjl01} and the
details will be given elsewhere~\cite{ddh01b}. We plot the renormalized $f_P
a^2$ in Fig. 10 as a function of $m_0 a$. We see that it rises sharply at small
quark mass.  Since the log form in Eq. (\ref{chi_log}) falls off slower at
larger quark mass, it gives a more stable fit covering a larger range of the
quark mass than the exponential form. Based on this observation, we fit $f_P$
in the log form

\begin{figure}[tb]
\includegraphics{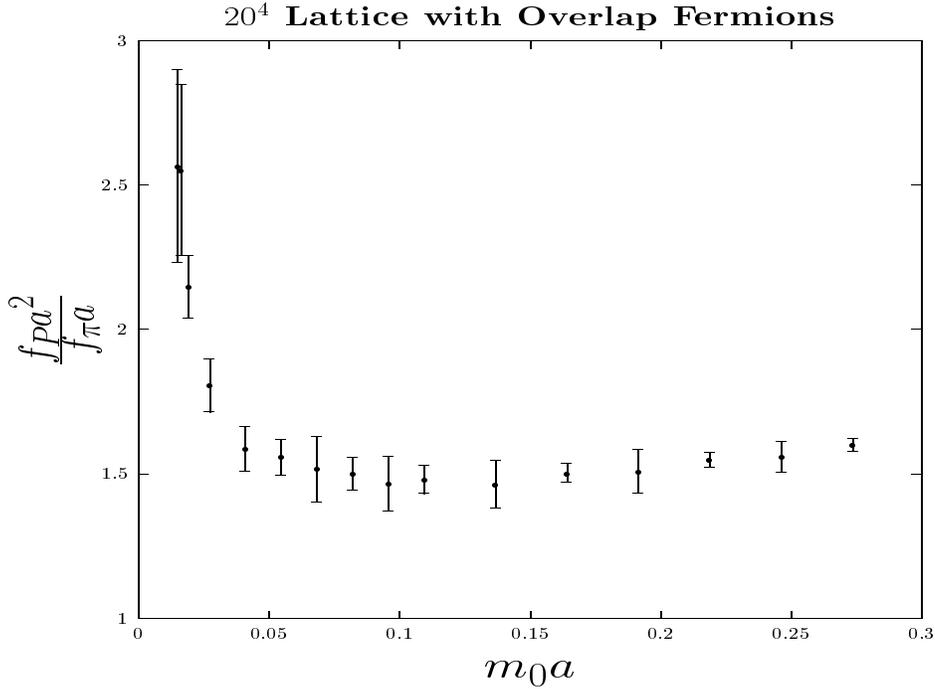}
\vspace{8cm}
\caption{Renormalized $f_Pa^2/f_{\pi}a$ vs $m_{\pi}^2 a^2$. The
solid line is a fit over the lowest 14 points.}
\end{figure}

\begin{equation}  \label{f_P_log}
f_P a^2 = \tilde{f_P} a^2 \{1 -\delta [\ln(Am_0 a/\Lambda_{\chi}^2 a^2) + 1]\} + B
m_0 a,
\end{equation}
where we input $A$ from the fit to $m_P^2$ in Eq. (\ref{chi_log}).  It turns
out that fits covering the whole range of 16 points have $\chi^2/DF$ greater
than unity mainly due to the fact the $f_P a^2$ of the two smallest quark
masses are too high. Since errors of these two points are large and they are
more susceptible to systematic errors from the boundary effect for such low
pion masses, we shall ignore them and fit Eq. (\ref{f_P_log}) with the other 14
points. Again, results with reasonably small errors and $\chi^2/DF$ less than
unity are reported in Table 6.

\begin{table}[htb]
\begin{center}
\caption{$\tilde{f_P} a^2$, $\delta$ and $\chi^2/DF$ as fitted
for $f_P a^2$ in Eq. (\ref{f_P_log}) are listed.}
\bigskip
\bigskip
\begin{tabular}{lllll}
\hline
$\Lambda_{\chi}$ (GeV) & $\tilde{f_P} a^2$  & $\delta$ & B & $\chi^2/DF$  \\
\hline
  0.6   &  0.083(2)  & 0.35(4) & 0.36(2) & 0.93   \\
  0.8   &  0.060(3)  & 0.48(7) & 0.36(2) & 0.93     \\
\hline
\end{tabular}
\end{center}
\end{table}

We see that $\delta$ obtained from $f_P a^2$ is consistent with those
determined from $m_{\pi}^2 a^2$. Combining all the fits from $m_{\pi}^2 a^2$
and $f_P a^2$, $\delta$ appears to span the range from 0.15 to 0.4 which is
consistent with that predicted from the quenched $\eta'$ loop in the chiral
perturbation theory in Eq. (\ref{delta})~\cite{sha92,bde01}, although it tends
to be on the high side.  We note that the general behavior of our data is close
to what are observed in Ref.~\cite{bde01,bbo01}, but our $\delta$ is about
three to six times larger than theirs.

To alleviate the apprehension that the apparent singular behavior of $f_P a^2$
may be caused by certain unknown boundary effect on the source in our lattice
with fixed boundary condition, we consider the ratio of $f_P$ in Eq.
(\ref{f_P_G_PP}) and $f_{\pi}$ in Eq. (\ref{f_pi})
\begin{equation}
f_P a^2/ f_{\pi}a = \lim_{t \longrightarrow \infty}\frac{Z_P \sqrt{2} m_P a
G_{\partial_4A_4P}(\vec{p}= 0, t)} {2 m_0a\, G_{A_4P}(\vec{p}= 0, t)}.
\end{equation}
This should cancel out the boundary effect on the source.  We plot the ratio
$f_P a^2/ f_{\pi}a$ as a function of $m_{\pi}^2 a^2$.  We see from Fig. 11 that
the singular behavior is still visible in $f_Pa^2/f_{\pi}a$ which confirms that
the chiral log singularity is indeed present in $f_P$.

\section{Summary}

To conclude, we have studied the chiral properties of the pseudoscalar meson on
a quenched lattice with overlap fermions. The lattice size is $20^4$ with
lattice spacing $a = 0.148\, {\rm fm}$ set by the pion decay constant
$f_{\pi}$. This gives a physical size of 3.0 fm which is about 4 times the
Compton wavelength of the lowest mass pion.

We first clarified the role of the zero modes in the pseudoscalar meson
propagator in association with the generalized Gell-Mann-Oakes-Renner
relation. We find that the zero mode contribution to the pseudoscalar meson
propagators extends to a fairly long distance in the time separation and it is
imperative to avoid it since it is a finite volume effect. Otherwise, one might
be led to wrong conclusions about the behavior of the pion mass and decay
constants near the chiral limit. For example, if one fits the pion mass by
choosing a time window which straddles over the kink in Fig. 2, one will find
that the pion mass squared does not approach zero at the chiral limit with a
linear extrapolation in the quark mass. Interpreting it as due to the quenched
chiral log, one may fit it with a chiral log form and obtain a positive
$\delta$ which can be misleading. In view of the fact that the quenched chiral
log fit of $m_{\pi}^2 a^2$ from data obtained from the shorter time range of
$G_{PP}(\vec{p}= 0, t)$ as plotted in Fig. 5 has a much smaller $\delta$ than
the present study and is consistent with that obtained in Ref.~\cite{bde01}, it
would be useful to clarify that the shifted real modes (the `would be' zero
modes) in the Wilson action are not responsible for the observed quenched
chiral logs in $m_{\pi}^2 a^2$ and $f_P a^2$ in this case.

Due to the chiral symmetry of the overlap fermion, we obtain the
non-perturbative renormalization constant $Z_A = 1.589(4)$ from the axial Ward
identity. We find it to be fairly independent of $m_0 a$ which is an indication
that the $O(a^2)$ error is small. The renormalized pion decay constant
$f_{\pi}$ has a positive slope with respect to $m_0 a$. With a small error
(1.6\%) and devoid of the complication of the quenched chiral log, $f_{\pi}$ is
an ideal physical observable to set the scale of the lattice.

We studied extensively the issue of quenched chiral log. We have clearly seen
the quenched chiral log both in the pseudoscalar meson masses with equal quark
masses and their pseudoscalar decay constants. Various fits give $\delta$ to be
in the range of 0.15 to 0.4 which is in accord with that estimated from the
double $\eta$ propagator approximation of the quenched $\eta'$ loop which
predicts it to be $\sim 0.18$.

We finally have a reliable tool in the overlap fermion to study the chiral
symmetry properties of hadrons at low energies including the quenched chiral
logs. One should go to different lattice spacings to study the continuum limit
in the future.

This work is partially supported by DOE Grants DE-FG05-84ER40154 and
DE-FG02-95ER40907.  We wish to acknowledge enlightening discussions and
exchanges with L. Giusti, M. Golterman, P. Hernandez, K. Jansen, D. Lin,
M. L\"{u}scher, R. Sommer, A. Soni, and C. Rebbi.

\end{document}